\documentclass[12pt,12pt,sort&compress]{article}

\usepackage[T1]{fontenc}
\usepackage[latin9]{inputenc}
\usepackage{amsmath}
\usepackage{amssymb}
\usepackage{graphicx}
\usepackage[numbers]{natbib}

\makeatletter

\newcommand{\lyxdot}{.}

\newcommand{\lyxaddress}[1]{
\par {\raggedright #1
\vspace{1.4em}
\noindent\par}
}

\@ifundefined{date}{}{\date{}}

\usepackage{indentfirst}
\usepackage{amsfonts}
\usepackage[T1]{fontenc}
\usepackage{ae,aecompl}
\usepackage{sidecap}
\usepackage[section]{placeins}
\usepackage{epsf}

\makeatother

\begin{document}

\title{Effect of interface phase transformations on diffusion and segregation
in high-angle grain boundaries}

\author{T. Frolov$^{1}$, S. V. Divinski$^{2}$, M. Asta$^{1}$ and Y. Mishin$^{3}$}

\maketitle

\lyxaddress{$^{1}$ Department of Materials Science and Engineering, University
of California, Berkeley, California 94720, USA}

\lyxaddress{$^{2}$ Institute of Materials Physics, University of Münster, Wilhelm-Klemm-Str.
10, D-48149 Münster, Germany}

\lyxaddress{$^{3}$ School of Physics, Astronomy and Computational Sciences,
George Mason University, Virginia, 22030, USA}
\begin{abstract}
Recent experimental measurements of Ag impurity diffusion in the $\Sigma5\,(310)$
grain boundary (GB) in Cu revealed an unusual non-Arrhenius behavior
suggestive of a possible structural transformation {[}Divinski \emph{et
al}., Phys.~Rev.~B 85, 144104 (2012){]}. On the other hand, atomistic
computer simulations have recently discovered phase transformations
in high-angle GBs in metals {[}Frolov \emph{et al.}, arXiv:1211.1756v2
(2013){]}. In this paper we report on atomistic simulations of Ag
diffusion and segregation in two different structural phases of the
Cu $\Sigma5\,(310)$ GB which transform to each other with temperature.
The obtained excellent agreement with the experimental data validates
the hypothesis that the unusual diffusion behavior seen in the experiment
was caused by a phase transformation. The simulations also predict
that the low-temperature GB phase exhibits a monolayer segregation
pattern while the high-temperature phase features a bilayer segregation.
Together, the simulations and experiment provide the first convincing
evidence for the existence of structural phase transformations in
high-angle metallic GBs and demonstrate the possibility of their detection
by GB diffusion measurements and atomistic simulations.
\end{abstract}
\textbf{\emph{Motivation.}} Structural transformations at grain boundaries
(GBs) are of fundamental interest as a case of phase transitions in
low-dimensional systems. In addition, GB phase transformations can
have a significant impact on microstructure, mechanical behavior and
transport properties of polycrystalline materials \citep{Balluffi95,Mishin2010a}.
A number of GB phases have been found in ceramic materials \citep{Baram08042011,Harmer08042011},
where they often appear in the form of intergranular thin films and
are referred to as \textquotedblleft{}complexions\textquotedblright{}
\citep{Dillon2007}. In metallic alloys, several phases with discrete
thickness have been observed, such as the segregated bilayer structure
believed to be responsible for the liquid-layer embrittlement effect
\citep{Luo23092011}. However, despite decades of research little
is known about GB phase transformations in single-component metals,
apart from the recently found dislocation pairing transition in low-angle
GBs composed of discrete dislocations \citep{Olmsted2011} and the
prediction of temperature-induced phase transformations in high-angle
GBs \citep{Frolov2013}. Presently, there is no direct experimental
evidence for structural transformations in metallic GBs. The main
reason lies in the extreme difficulty of GB structure characterization
by high-resolution transmission electron microscopy (HRTEM) at high
temperatures \citep{Merkle1987,Duscher04}.

\begin{figure}
\noindent \begin{centering}
\includegraphics[scale=0.7]{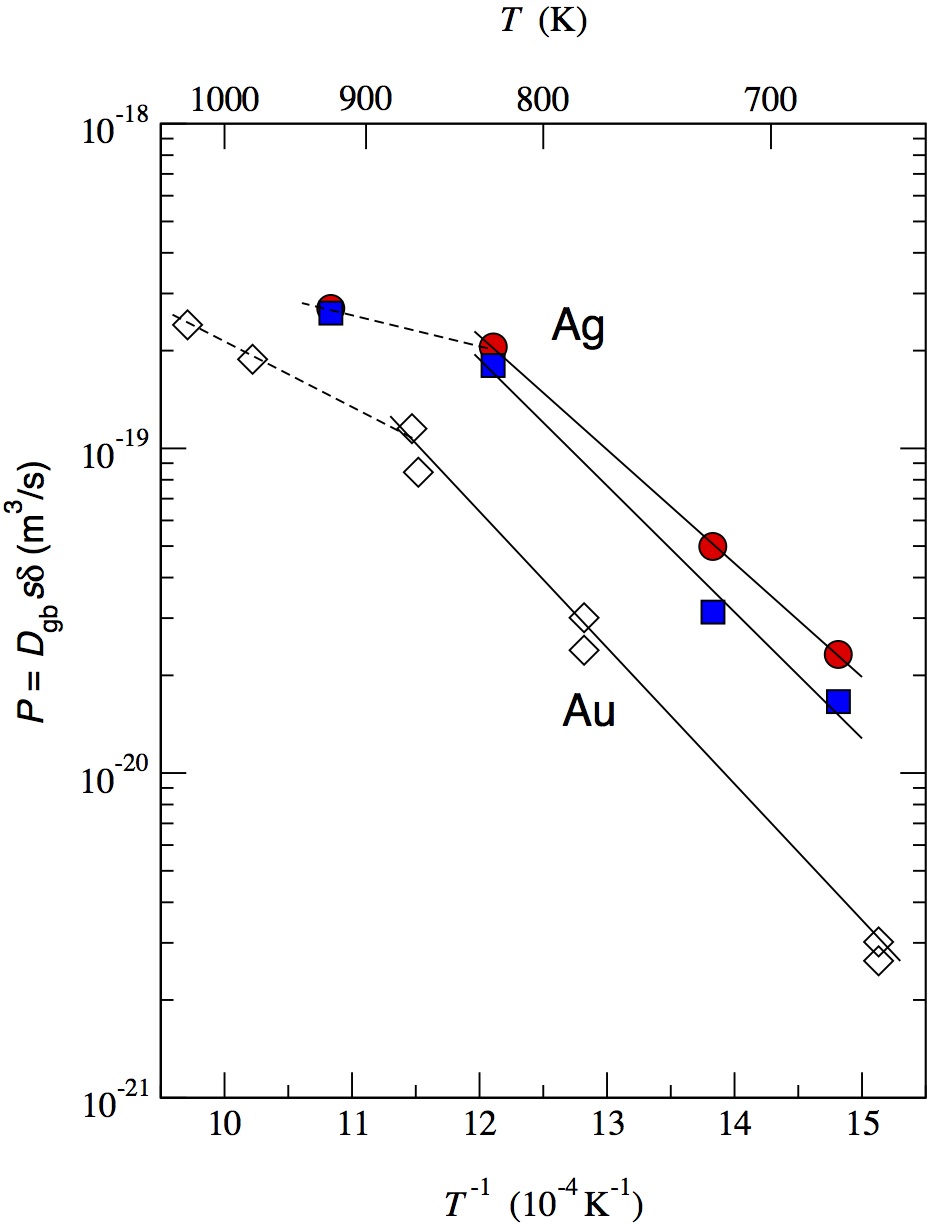}
\par\end{centering}

\caption{Arrhenius diagram of impurity diffusion of Ag \citep{Divinski2012}
and Au \citep{Budke1999} the Cu $\Sigma5\,(310)$ GB obtained by
radio-tracer measurements on bicrystals. For Ag diffusion, the circles
and squares are for diffusion parallel and normal to the tilt axis,
respectively. Note the break in the slope of both plots at high temperatures.
\label{fig:Arrhenius}}

\end{figure}

Several indirect methods have been applied to probe the structural
state of GBs, including GB diffusion \citep{Kaur95}, GB mobility
\citep{Gottstein} and GB sliding resistance \citep{Watanabe84b}.
Recently, impurity diffusion of Ag in a Cu bicrystal with a well-characterized
$\Sigma5\,(310)$ GB has been measured using a radioactive isotope
of Ag ($\Sigma$ being the reciprocal density of coincidence sites
and (310) the GB plane) \citep{Divinski2012}. At high temperatures,
the GB diffusivity was characterized by the diffusion flux $P=D_{gb}s\delta$,
where $D_{gb}$ is the GB diffusion coefficient, $s$ is the impurity
segregation factor and $\delta$ the thickness of the GB core considered
as a uniform layer. The Arrhenius plot shown in Fig.~\ref{fig:Arrhenius}
reveals a distinct break in the slope at temperatures around 800 to
850 K. This observation correlates with the similar break found previously
for Au diffusion in a similar GB \citep{Budke1999}. Such breaks in
the slopes of Arrhenius plots usually signify an abrupt change in
the diffusion mechanism \citep{Kaur95,Mishin99f}. On this ground,
it was suggested \citep{Divinski2012,Budke1999} that the $\Sigma5\,(310)$
GB undergoes a structural phase transformation at about 800-850 K.
While a plausible explanation, its confirmation by direct experimental
observations is hampered by the mentioned difficulties in HRTEM characterization
of GBs at high temperatures. 

\begin{figure}
\noindent \begin{centering}
\includegraphics[scale=0.35]{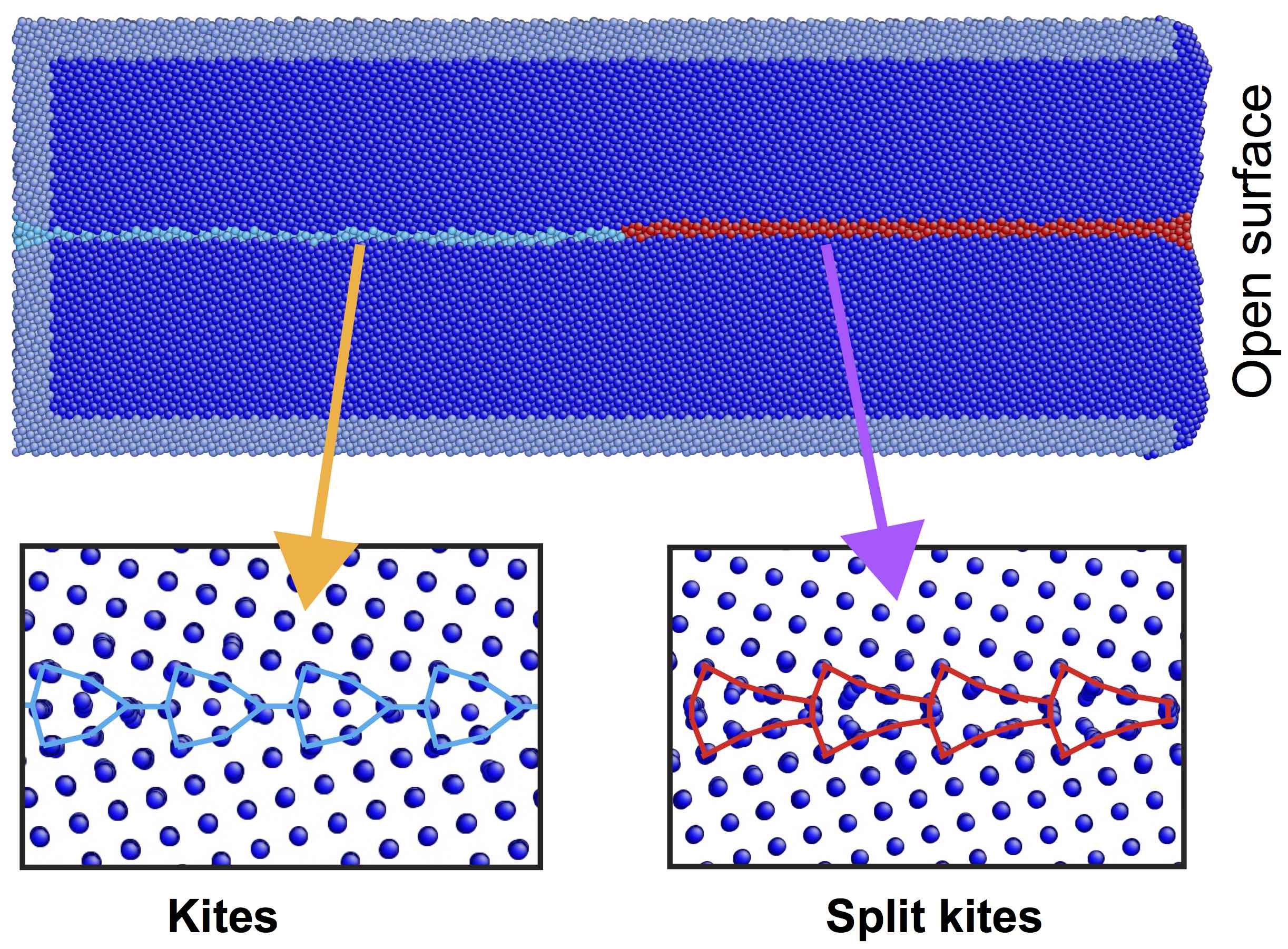}
\par\end{centering}

\caption{Phase transformation in the Cu $\Sigma5\,(310)$ GB at the temperature
of 1000 K \citep{Frolov2013}. The initial structure of the boundary
was composed of kite-shape structural units (light blue). A new phase
composed of split-kite structural units (red) grows into the GB and
eventually penetrates all through the sample. This transformation
is enabled by supply of atoms from the surface. \label{fig:Phase-transformation}}

\end{figure}

Much of the current knowledge about GB structures comes from atomistic
computer simulations \citep{Balluffi95,Mishin2010a}. Many previous
simulations have shown a behavior where GBs exhibit structural disorder
at high temperatures and eventually melt by either turning into a
continuously growing liquid film or triggering bulk melting \citep{Balluffi95,Mishin2010a,Divinski2012,Suzuki05a,Frolov-2011}.
However, until the recent report \citep{Frolov2013} there was no
compelling simulation evidence for transformations between different
\emph{ordered} structures in high-angle metallic GBs. Observation
of such transformations had been precluded by inadequate simulation
methodology which prohibited variations in atomic density in the GB
core. When such variations were allowed, several alternate structures
and first-order phase transitions between them were found in the $\Sigma5\,(210)$
and $\Sigma5\,(310)$ GBs in FCC metals \citep{Frolov2013}. For the
Cu $\Sigma5\,(310)$ GB, a transformation between two structures was
found, with split-kite structural units stable above 800 K and kite
units stable at lower temperatures (Fig.~\ref{fig:Phase-transformation}).
The kite structure had been known previously, whereas the split-kite
structure is new and was first discovered in \citep{Frolov2013}.

The finding of this phase transformation \citep{Frolov2013} and the
correlation between its temperature and the experimental temperature
of the change in the diffusion activation energy \citep{Divinski2012}
lend more credence to the hypothesis that the observed diffusion behavior
was caused by a GB phase transformation. However, a convincing proof
requires a calculation of the diffusivity in impurity-segregated boundaries
in the vicinity of the transformation temperature and demonstration
that it indeed reproduces the experiment. It is the goal of this paper
to conduct such calculations and validate the proposed hypothesis.
We also wish to understand the effect of the transformation on Ag
segregation and the role of this segregation in the diffusion behavior.

\textbf{\textit{Methodology.}} The molecular dynamics (MD) simulations
were performed in the canonical (NVT) ensemble and employed the LAMMPS
code \citep{Plimpton95} and the embedded-atom potential for copper
\citep{Mishin01}. The simulation block with a plane $\Sigma5\,(310)$
GB with the standard kite structure was created by the usual geometric
construction \citep{Suzuki03a} and contained around 47,000 atoms.
The block had the approximate dimensions of 6.6$\times$12.5$\times$7.00
nm$^{3}$ with periodic boundary conditions parallel to the GB plane
$x$-$z$ and free surfaces in the $y$ direction normal to the GB
plane. The split-kite GB structure was first created by an isothermal
800 K anneal of a GB terminated at free surfaces. The central part
of this block was then carved out to create a new simulation block
with the same dimensions and the same boundary conditions as for the
standard kite structure. The obtained GBs (one with kites and the
other with split kites) were disconnected from sinks and sources of
atoms and unable to vary their density. Consequently, their structures
remained unaltered during the subsequent anneals. Using these simulation
blocks, diffusivities of the two GB structures could be studied separately
over the same temperature interval.

For each GB structure, a series of isothermal anneals was performed
at temperatures from 750 to 1000 K. The simulation time $t$ increased
from 10 ns at 1000 K to 70 ns at 750 K. Multiple snapshots saved during
the simulations were used to analyze the trajectories of atoms diffusing
within the GB core. The GB self-diffusion coefficient was computed
from the Einstein relation $D_{gb}^{||}=\left\langle x^{2}\right\rangle /2t$
for the direction parallel to the tilt axis and $D_{gb}^{\bot}=\left\langle z^{2}\right\rangle /2t$
for the perpendicular direction. To validate this methodology, we
verified that the mean-squared displacements indeed increased linearly
with time as prescribed by the Einstein equation.

While the obtained self-diffusion coefficients are useful as a guide
for future research (see discussion below), the experimental measurements
\citep{Divinski2012} were made for Ag impurity diffusion. The generalized
Fisher model \citep{Kaur95} employed in \citep{Divinski2012} for
the processing and interpretation of the experimental concentration
curves assumes that the impurity atoms segregate to the GB and remain
in thermodynamic equilibrium with adjacent lattice regions. It was
therefore necessary to create equilibrium segregation of Ag in the
simulated models. To this end, we used the semi-grand canonical Monte
Carlo method \citep{FrenkelS02} with Cu-Ag interactions modeled with
the embedded-atom potential \citep{Williams06}. At each temperature
$T$, the desired equilibrium chemical composition $c$ inside the
grains was established by adjusting the imposed chemical potential
difference $\Delta\mu$ between Ag and Cu. The lattice constant was
chosen to eliminate mechanical stresses inside the grains. At each
$T$ and $\Delta\mu$, an equilibration anneal was performed followed
by a production run comprising $4\times10^{5}$ to $6\times10^{5}$
Monte Carlo steps per atom. Several dilute grain compositions ranging
from 0.001 to 0.1 atomic percent of Ag were implemented at each temperature.
At the temperature of 800 K, a more detailed calculation of the segregation
isotherm was performed with $c$ varying from 0.001\% to 0.6\%. 

The blocks with the equilibrium Ag segregation corresponding to the
temperatures between 750 and 1000 K and the same grain composition
$c=0.1$\% were taken as initial configurations for MD simulations
of Ag impurity diffusion. As for self-diffusion, $D_{gb}$ was extracted
from mean-squared atomic displacements of Ag atoms in the GB core.
Since the number of such atoms was significantly less than for self-diffusion,
longer MD simulation times ranging from 50 to 70 ns were implemented.

\begin{figure}
\noindent \begin{centering}
\includegraphics[width=0.9\textwidth]{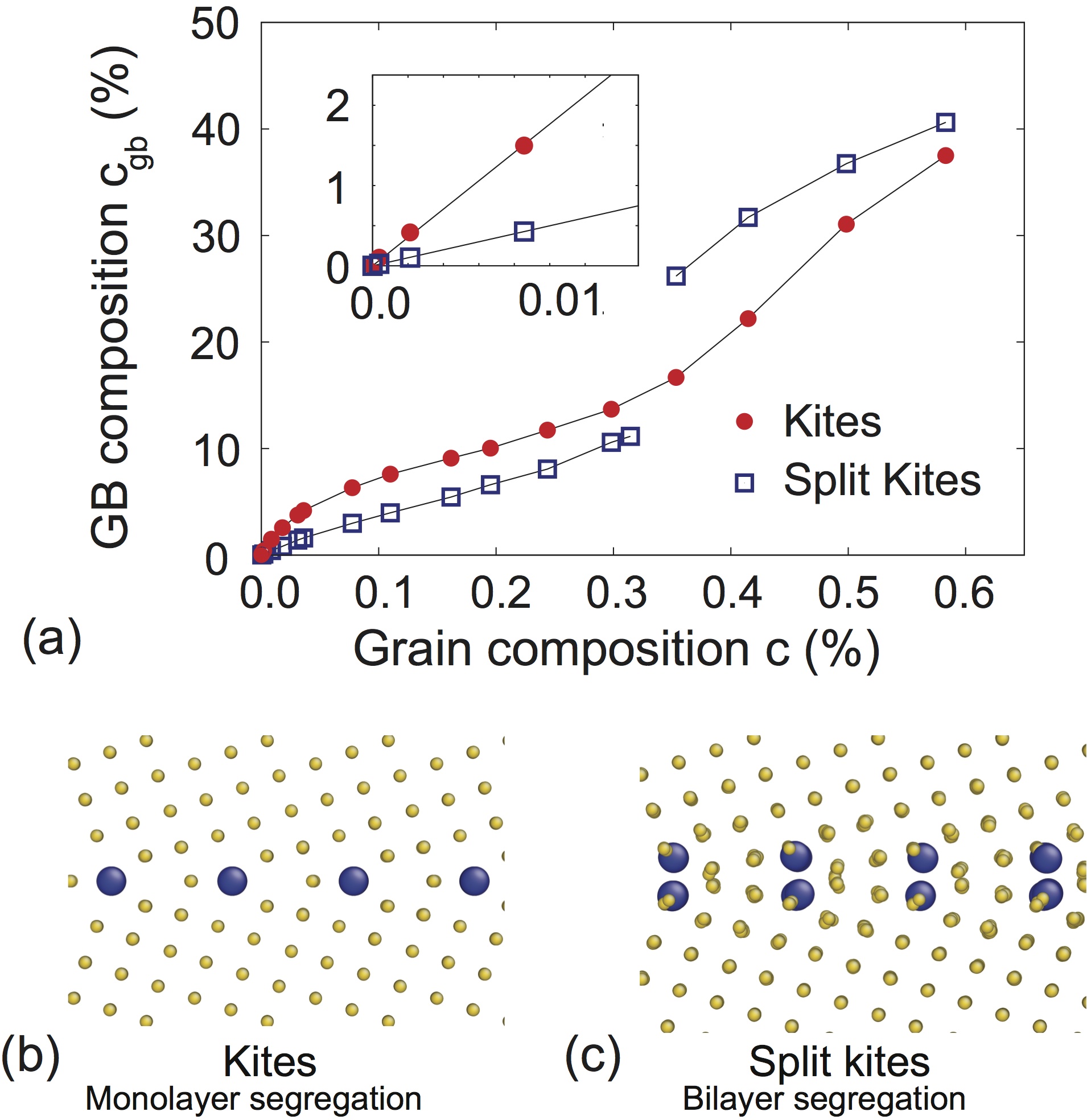}
\par\end{centering}

\caption{(a) Isotherm of Ag segregation at the Cu $\Sigma5\,(310)$ GB at the
temperature of 800 K. The inset shows the low-concentration part used
for the calculation of the dilute segregation factor. The images show
the segregation patterns for the kite (b) and split-kite (c) GB structures.
For clarity the atomic structures were quenched to 0 K to eliminate
the thermal noise due to atomic vibrations. \label{fig:Isotherm}}
\end{figure}

\textbf{\textit{Results.}} Fig.~\ref{fig:Isotherm}(a) shows the
segregation isotherm at 800 K. The GB concentration $c_{gb}$ was
obtained by computing the amount of Ag segregation at a constant total
number of atoms \citep{Frolov2012b} and assuming the GB thickness
$\delta=$ 0.5 nm. The segregation factor $s=c_{gb}/c$ appearing
in the generalized Fisher model \citep{Kaur95} must be computed in
the dilute regime where it is independent of the grain composition.
The dilute regime was also implemented in the diffusion measurements
\citep{Divinski2012}, where the radio-tracer Ag atoms only served
as a probe of the GB structure. The dilute segregation factor was
obtained by a linear fit through the first 3 points near the origin
of the plot. Similarly, the segregation factors at all other temperatures
were calculated using the first 2-3 points where the isotherm was
linear (Henry-type \citep{Balluffi95,Kaur95}). 

It is interesting to examine the segregation behavior beyond the linear
regime. As evident from Fig.~\ref{fig:Isotherm}(a), the segregation
becomes non-linear and qualitatively consistent with the McLean isotherm
\citep{Balluffi95} between 0.05\% and 0.1\%. At higher concentrations,
however, $c_{gb}$ does not saturate as expected from the McLean isotherm
but continues to increase. Furthermore, at about 0.2\% the isotherm
goes through an inflection point and the increase in $c_{gb}$ accelerates.
This behavior is consistent with the experimental isotherms reconstructed
from GB diffusion data \citep{Divinski2005} and is usually indicative
of multiple types of segregation sites and/or segregation to multiple
layers. Note that the segregation to the standard kite structure is
significantly stronger than to the split kites. In addition, the two
GB structures exhibit different segregation patterns as illustrated
in Fig.~\ref{fig:Isotherm}(b,c). In the case of standard kites,
the Ag atoms tend to occupy the tip of the kite, whereas in the split
kites they segregate to corners of the kites, creating a bilayer segregation
pattern. It should be emphasized that neither the monolayer nor the
bilayer are completely filled with Ag atoms. In fact, the total number
of atoms segregated into the bilayer is less than for the monolayer.
Based on the segregation isotherm one can predict that if the GB structure
at a given temperature transforms from standard kites to split kites,
the equilibrium Ag segregation must abruptly drop while the segregation
type must change from a monolayer to a bilayer. As already mentioned,
changes in the segregation type from a monolayer to a bilayer (and
possibly three or more layers) were observed experimentally in metallic
alloys \citep{Luo23092011}.

It can also be observed that at $c\approx0.3$\% the segregation to
the split-kite structure abruptly increases. The new GB structure
which forms above 0.3\% is less ordered and does not follow the bilayer
pattern. This segregation jump is suggestive of a segregation-induced
structural transformation reminiscent of the one found previously
in twist GBs \citep{Seidman96}. However, this simulation setup does
not permit a conclusive identification of the new phase because the
GB is not free to adjust its total atomic density. A more comprehensive
study would employ a GB terminated at free surfaces and is beyond
the scope of this paper. 

\begin{figure}
\noindent \begin{centering}
\textbf{(a)} \enskip{}\includegraphics[scale=0.45]{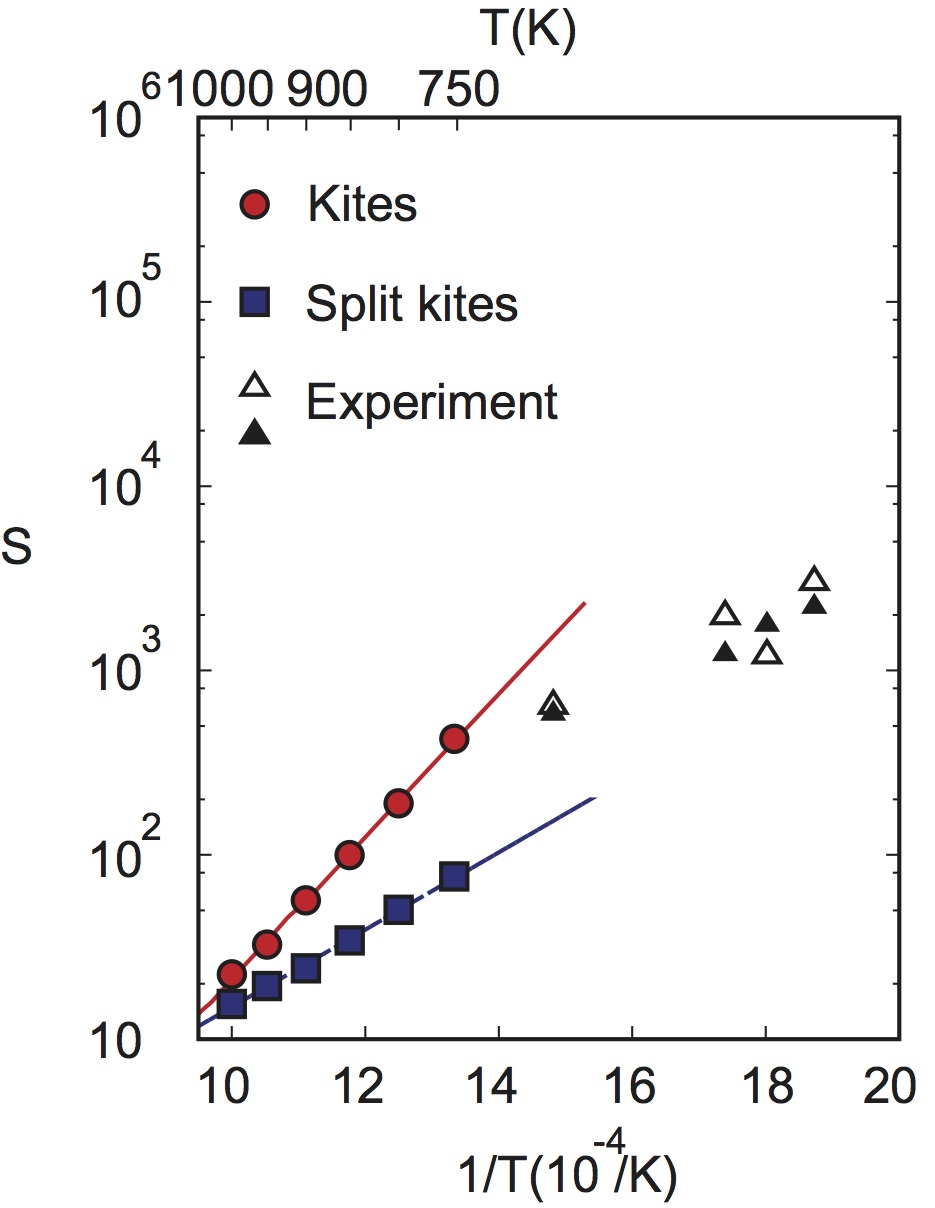}
\par\end{centering}

\noindent \begin{centering}
\textbf{(b)} \enskip{}\includegraphics[scale=0.45]{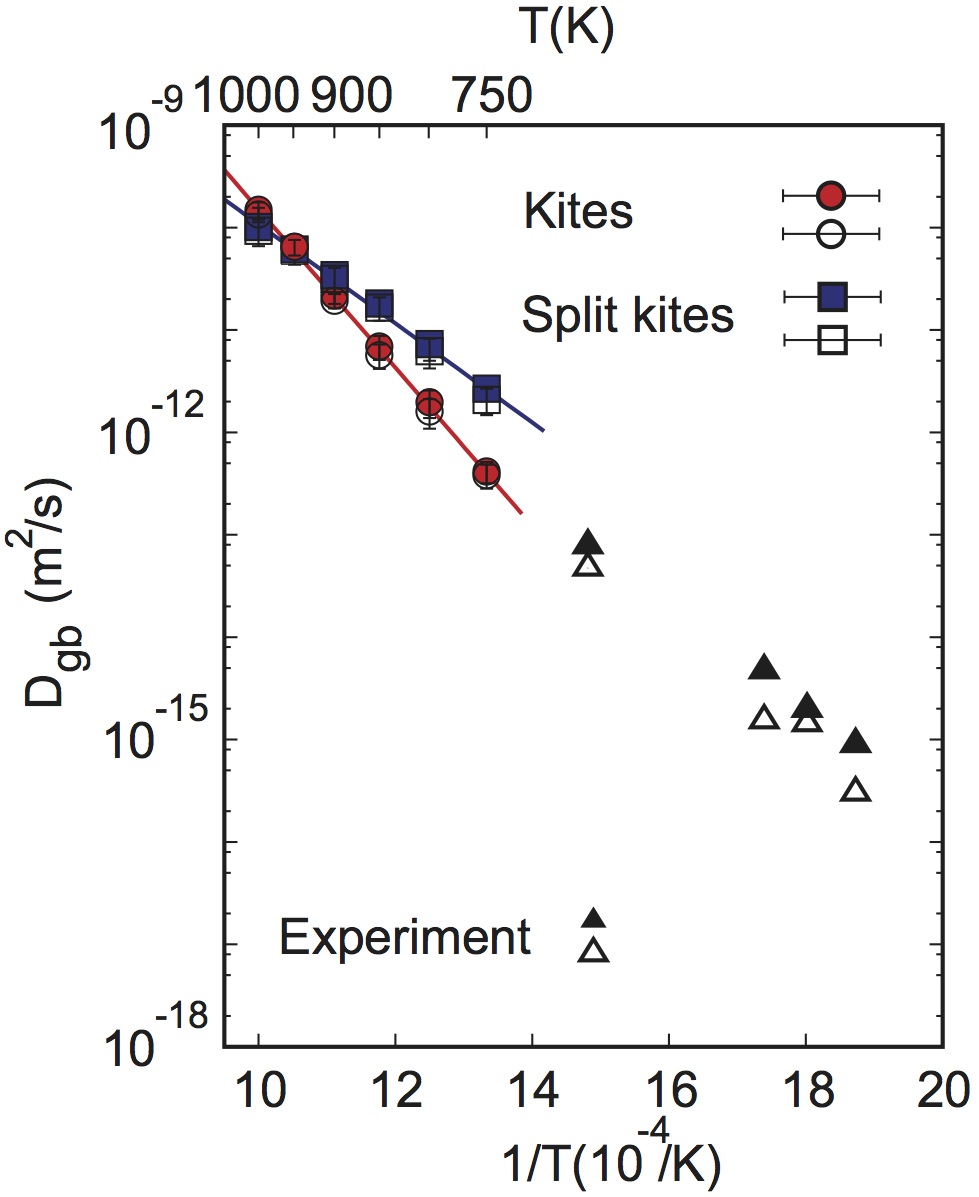}
\par\end{centering}

\caption{Arrhenius diagrams for (a) dilute segregation factor $s$ and (b)
diffusion coefficient of Ag in the Cu $\Sigma5\,(310)$ GB. The experimental
data \citep{Divinski2012} is shown for comparison. The experimental
segregation factors were back-calculated from diffusion data. Diffusion
parallel and normal to the tilt axis is represented by filled and
open symbols, respectively.\label{fig:Arrhenius-diagrams}}

\end{figure}

The Arrhenius diagram in Fig.~\ref{fig:Arrhenius-diagrams}(a) summarizes
the dilute segregation factors for the kite and split-kite structures.
Both structures accurately follow the Arrhenius law but with significantly
different slopes. The effective segregation energy extracted from
the slopes is 74 kJ/mole for the kite structure and 40 kJ/mole for
the split-kite structure. Included are also the experimental segregation
factors reported in \citep{Divinski2012}. It should be noted that
they were not obtained by direct chemical characterization of the
GB but rather back-calculated from the diffusivities measured in different
kinetic regimes. Nevertheless, given that the segregation factors
span nearly two orders of magnitude, the agreement between the simulations
and experiment can be considered favorable.

Similarly, the diffusion coefficients of Ag in the two GB structures
accurately follow the Arrhenius law {[}Fig.~\ref{fig:Arrhenius-diagrams}(b){]}
with the activation energies 149 kJ/mole for the kite structure and
93 kJ/mole for the split-kite structure (averaged over the two diffusion
directions). The diffusion coefficient is lower for the split-kite
structure at high temperatures and for the standard kite structure
at low temperatures. The temperature of the cross-over between the
two diffusivities correlates well with the experiment (cf.~Fig.~\ref{fig:Arrhenius}).
Furthermore, extrapolation of the kite diffusivity to lower temperatures
gives good agreement with the experimental $D_{gb}$ values measured
in the type-C kinetic regime \citep{Divinski2012}.

\begin{figure}
\noindent \begin{centering}
\includegraphics[height=0.5\textheight]{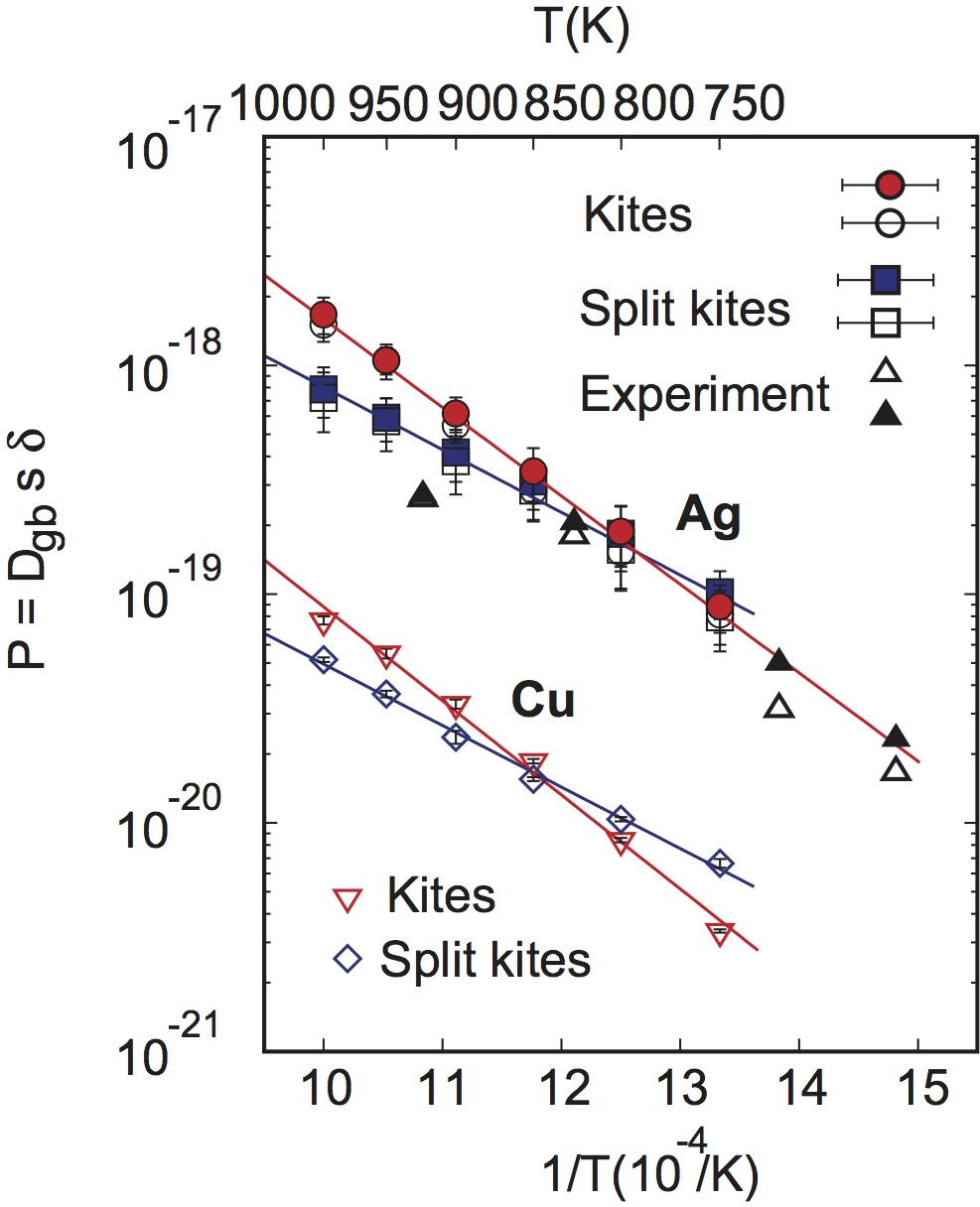}
\par\end{centering}

\caption{Arrhenius diagram of the diffusion flux for Ag impurity diffusion
(upper curves) and Cu self-diffusion (lower curves) in the Cu $\Sigma5\,(310)$
GB. For Ag diffusion, the experimental data \citep{Divinski2012}
is shown for comparison. Diffusion parallel and normal to the tilt
axis is represented by filled and open symbols, respectively.\label{fig:Flux}}

\end{figure}

Putting together the diffusion and segregation data, Fig.~\ref{fig:Flux}
shows the Arrhenius diagram of the GB flux $P=D_{gb}s\delta$. The
first notable feature of this plot is the overall close agreement
between the experimental measurements \citep{Divinski2012} and independent
simulation data. For both GB phases, the computed fluxes $P$ accurately
follow the Arrhenius law with the activation energies 74 kJ/mole for
the kite phase and 52 kJ/mole for the split-kite phase (averaged over
both directions). For the kite phase stable at low temperatures, this
activation energy compares reasonably well with the experimental values
(67 to 75 kJ/mole \citep{Divinski2012}) extracted from the low-temperature
portion of the experimental Arrhenius diagram. 

The cross-over between the diffusivities $P$ of the two GB phases
occurs at the temperature of  800 K, which is in excellent agreement
with the experiment. This cross-over creates the characteristic break
in the slope of the Arrhenius plot which is strikingly similar to
the one reported in \citep{Divinski2012}. This close agreement strongly
suggests that the non-Arrhenius diffusion behavior found in the experiment
\citep{Divinski2012} was caused by the structural phase transformation
in this boundary.

\textbf{\textit{Conclusions.}} One general conclusion from this work
is that atomistic simulations are presently capable of predicting
GB diffusion and segregation in good agreement with experiment. It
has also been demonstrated that our simulation methodology permits
diffusion and segregation calculations for \emph{individual} GB phases.
This opens the door for future investigations of the effect of GB
phase transformations on thermodynamic (segregation) and kinetic (diffusion)
properties of high-angle GBs in metals by means of atomistic modeling.
For the particular Cu $\Sigma5\,(310)$ GB studied here, the simulations
predict that the structural transformation must cause a break in the
activation energy of Ag impurity diffusion at about 800 K, with a
lower activation energy above this temperature. This prediction is
in excellent agreement with the recent experimental study \citep{Divinski2012}.
This agreement confirms that the non-Arrhenius diffusion behavior
reported in \citep{Divinski2012} indeed constitutes an experimental
observation of a structural phase transformation in a high-angle GB.

Our simulations predict that the phase transformation must cause a
similar break in the activation energy for Cu self-diffusion (Fig.~\ref{fig:Flux}).
This suggests that the non-Arrhenius behavior found for Ag diffusion
in Cu \citep{Divinski2012} was primarily caused by the effect of
the phase transformation on the diffusion coefficient $D_{gb}$ rather
than on segregation. Nevertheless, segregation also plays an important
role as it affects the activation energies of the measured diffusion
fluxes $P$ in the two phases. Direct segregation measurements may
provide additional information on the GB transformation. Our study
shows that atomistic simulations are capable of reproducing complex
segregation patterns, such as the bilayer, found experimentally in
metallic alloys \citep{Luo23092011} and ceramics \citep{Baram08042011,Harmer08042011}.
This work motivates future measurements of Cu self-diffusion in the
$\Sigma5\,(310)$ GB to test the predictions shown in Fig.~\ref{fig:Flux}.
Cu self-diffusion measurements can employ the radioactive isotopes
$^{64}$Cu or $^{67}$Cu and are extremely challenging \citep{Surholt94,Surholt97a}.
Nevertheless, such measurements could be conducted in the near future
using bicrystalline samples as it was done in \citep{Divinski2012}.

T.F. was supported by a post-doctoral fellowship from the Miller Institute
for Basic Research in Science at University of California, Berkeley.
Y.M. was supported by the U.S. Department of Energy, the Physical
Behavior of Materials Program, through Grant No. DE-FG02-01ER45871.
SVD was supported by the German Science Foundation through Grant No.
DI 1419/3-2. 

\newpage{}\clearpage{}

\end{document}